\documentclass[twocolumn,showpacs,preprintnumbers,amsmath,amssymb]{revtex4}

\usepackage{graphics,epsfig,graphicx}
\usepackage{subfig}
\usepackage{color}

\newcommand {\mrm} {\mathrm}

\begin{document}

\title{Polarization-correlated photon pairs from a single ion}

\author{F. Rohde, J. Huwer, N. Piro, M. Almendros, C.
Schuck, F. Dubin and J. Eschner} \affiliation{ICFO-Institut de
Ci\`encies Fot\`oniques, Mediterranean Technology Park, E-08860
Castelldefels (Barcelona), Spain}
\date{\today}

\begin{abstract}
In the fluorescence light of a single atom, the probability for
emission of a photon with certain polarization depends on the
polarization of the photon emitted immediately before it. Here
correlations of such kind are investigated with a single trapped
calcium ion by means of second order correlation functions. A
theoretical model is developed and fitted to the experimental
data, which show 91\% probability for the emission of
polarization-correlated photon pairs within 24 ns.
\end{abstract}

\maketitle

\section{Introduction}

One of the most relevant tools for characterizing the light field
emitted by a quantum optical system is its intensity correlation
function $g^{(2)}(\tau)$, the most prominent example being the
observation of antibunching
\cite{Kimble1977PRLv39p691,Diedrich1987PRLv58p203} in the
fluorescence of a single atom. Intensity correlation functions of
the fluorescence light of a single ion excited by two light fields
have been investigated before. In \cite{Schubert1992PRLv68p3016}
the pair correlation conditioned on the wavelength of the photons
scattered by a $\mrm{Ba^+}$ ion (eight-level structure) reveal the
transient internal dynamics of the ion, which is characterized by
optical pumping and the excitation of Raman coherence. In a
detailed description of these experiments
\cite{Schubert1995PRAv52p2994}, it is shown that measuring the
correlation function for only one polarization of one of the two
light fields by adding a polarization filter projects the ion in a
coherent superposition of its energy eigenstates. This may be
considered a first signature of atom-photon entanglement
\cite{PET}.

Other investigations of the resonance fluorescence of coherently
driven single atoms by means of $g^{(2)}(\tau)$ comprise, e.g., a
theoretical analysis of the effect of the quantized motion of a
two-level atom in a harmonic trap on the second order correlation
function \cite{Jakob1999PRAv59p2111}, experimental measurements of
photon correlations revealing single-atom dynamics in a
magneto-otical trap
\cite{Gomer1998PRAv58p1657,Gomer1998APBv67p689} and atomic
transport in an optical lattice \cite{Jurczak1996PRLv77p1727}, the
theoretical demonstration of nonclassical correlations in the
radiation of two atoms \cite{Skornia2001PRAv64p63801}, i.e.
bunched and anti-bunched light is emitted in different spatial
directions, and the study of intensity-intensity correlations from
a three-level atom damped by a broadband squeezed vacuum
\cite{Carreno2004JoOBv6p315}.

Another application of photon-photon correlations in the resonance
fluorescence from single atoms is an experiment where, using a
self homodyning configuration, the second order correlation
function was used to characterize the secular motion of a trapped
ion \cite{Rotter2008NJPv10p43011}. The experiment revealed the
dynamics of both internal and external degrees of freedom of the
ion's wave function, from nanosecond to millisecond timescales, in
a single measurement run.

Polarization correlations of subsequent photons enitted from a
bichromatically driven four-level atom have been predicted
theoretically \cite{Jakob2002JOBv4p308}. In this proposal a
four-level atom ($\mrm{J=1/2}$ to $\mrm{J=1/2}$) with degenerate
Zeeman sub-states is coupled to a light field consisting of two
components which are symmetrically detuned from the resonance
frequency of the atomic transition. In contrast, in the experiment
presented in this paper, the Zeeman degeneracy is lifted and the
corresponding atomic transition (four-level system) is driven
monochromatically. Furthermore, the effect of the coupling of the
excited states to another manifold ($\mrm{J=3/2}$) of four Zeeman
sub-levels by a second light field is taken into account.


In the context of quantum optical information technologies, many
entanglement protocols strongly rely on a projective measurement
\cite{Cabrillo1999PRAv59p1025,Browne2003PRLv91p67901,Feng2003PRLv90p217902,Duan2003PRLv90p253601,Simon2003PRLv91p110405},
which is also the origin of the effect of
antibunching. 
In fact, the degree of antibunching is a benchmark for single
quantum emitters or ideal single photon sources that are suitable
for quantum networking and communication.

Beyond proving that a source is a pure single quantum emitter,
correlation functions are used to characterize and develop
applications for quantum communication. In
\cite{Dubin2007PRLv99p183001}, for example, the resonance
fluorescence from a continuously excited single ion is split and
recombined on a beam splitter with a relative delay, thus creating
an effective two-photon source. Here the measured correlation
function reveals the quality of the mode matching at the beam
splitter. Measuring and controlling the degree of second order
correlation of a photon source is therefore a very important step
in designing quantum optical tools in quantum information
processing. This engineering of correlation functions is of
special interest for quantum networks where single photons mediate
information between nodes of single atoms
\cite{Zoller2005EPJDv36p203,Briegel1998PRLv81p5932}.

In this context, we present a polarization selective measurement
of the correlation function $g^{(2)}(\tau)$ of fluorescence
photons from a single $^{40}\mrm{Ca^+}$ ion, which is continuously
laser excited on the $\mrm{4^2S_{1/2}}$ to $\mrm{4^2P_{1/2}}$
transition. In particular, the correlation function of emitted
$\sigma^-$ and $\sigma^+$ polarized photons \footnote{Note that we
identify the photon polarization ($\sigma^-$ or $\sigma^+$) by the
transition on which the photon has been emitted.} was measured
conditioned on the previous detection of a $\sigma^-$ photon, and
it is shown that the system can be tailored such that the
polarization of a photon depends strongly on the polarization of a
previous one.

\section{Model} \label{Model}
\begin{figure}
\begin{center}
\includegraphics[width=\columnwidth]{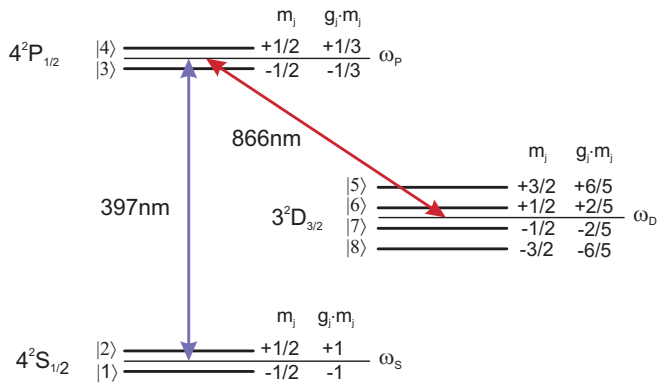}
\caption{Level scheme of $\mrm{^{40}Ca^+}$ for a non zero magnetic
field $\vec{\mrm{B}}$. The $\mrm{^2S_{1/2}}$, $\mrm{^2P_{1/2}}$
and $\mrm{^2D_{3/2}}$ states split into eight levels according to
their magnetic quantum number $m_j$ and the Land\'e factors $g_j$.
The splitting shown is not to scale.}
\label{Termschema_Ca_8levels}
\end{center}
\end{figure}

Figure \ref{Termschema_Ca_8levels} shows the relevant levels of
the $^{40}\mrm{Ca^+}$ ion. Continuous excitation with laser light
at 397~nm and 866~nm generates continuous resonance fluorescence.
We consider the case of observation along the direction of the
magnetic field and write $g^{(2)}(\tau)$ in terms of the
transition operators for the $\sigma$-transitions from
$\mrm{S_{1/2}}$ to $\mrm{P_{1/2}}$ (see figure
\ref{Termschema_Ca_8levels} for the numbering of the levels),
\begin{eqnarray}
\hat{\sigma}_1&=& |1\rangle \langle 4|, \hspace{1cm}
\hat{\sigma}_1^{\dagger}= |4\rangle \langle 1|,\label{sigma8level_a}\\
\hat{\sigma}_2&=& |2\rangle \langle 3|, \hspace{1cm}
\hat{\sigma}_2^{\dagger}= |3\rangle \langle 2|.
\label{sigma8level}
\end{eqnarray}
If we consider the general case without polarization selective
detection, the second order correlation function reads
\cite{Loudon2000}
\begin{equation}
g^{(2)}(\tau)=\frac{\sum_{i,j=1}^{2}  \langle
\hat{\sigma}^{\dagger}_{i}(t)\hat{\sigma}^{\dagger}_{j}(t+\tau)\hat{\sigma}_{j}(t+\tau)\hat{\sigma}_{i}(t)\rangle}{\langle
\hat{\sigma}^{\dagger}_{1}(t)\hat{\sigma}_{1}(t) +
\hat{\sigma}^{\dagger}_{2}(t)\hat{\sigma}_{2}(t)\rangle^2}.
\label{g2pi}
\end{equation}
Using the quantum regression theorem \cite{Mandel1995} it can be
shown that for all initial conditions and at steady state, the
second order correlation function is related to the excited state
populations by
\begin{equation}
g^{(2)}(\tau)=\frac{\rho_{33}(\tau)+\rho_{44}(\tau)}{\rho_{33}(\infty)+\rho_{44}(\infty)},
\label{g2rho3344}
\end{equation}
where $\rho_{lm}(\tau)$ with $l,m \in [1,8]$ are the matrix
elements of the density operator representing populations and
coherences of the states $|l\rangle$ and $|m\rangle$ at time
$\tau$ after the detection of a first photon. Equation
\ref{g2rho3344} allows to predict $g^{(2)}$ by solving the optical
Bloch equations. The second order correlation function
$g^{(2)}(\tau)$ of the 397~nm fluorescence light of a single
$^{40}\mrm{Ca^+}$ ion is thus proportional to the population of
the ion's excited state $\mrm{P_{1/2}}$ at time $\tau$ whereby the
initial state at $\tau=0$ is determined by the first measured
photon.

\begin{figure*}
\begin{center}
\subfloat[Projection]{\label{Termschema_Ca_4levels_cond_g2_prodject}\includegraphics[width=0.650\columnwidth]
{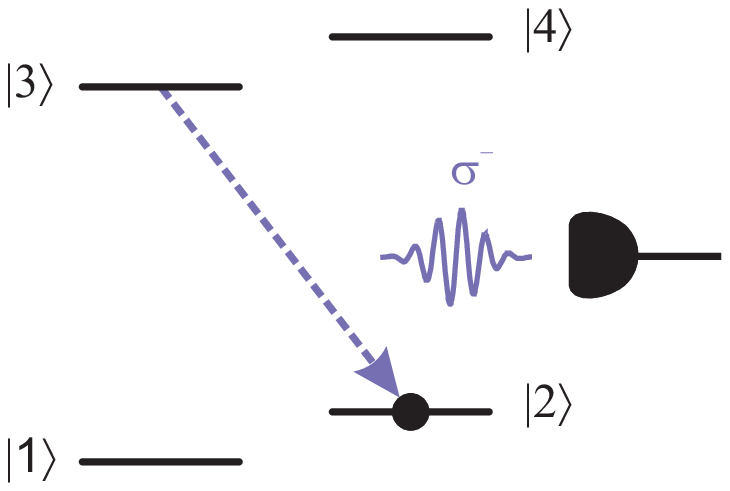}}
 \subfloat[$\sigma^{-}$ polarized photon]
   {\label{Termschema_Ca_4levels_cond_g2_sigma-}\includegraphics[width=0.650\columnwidth]{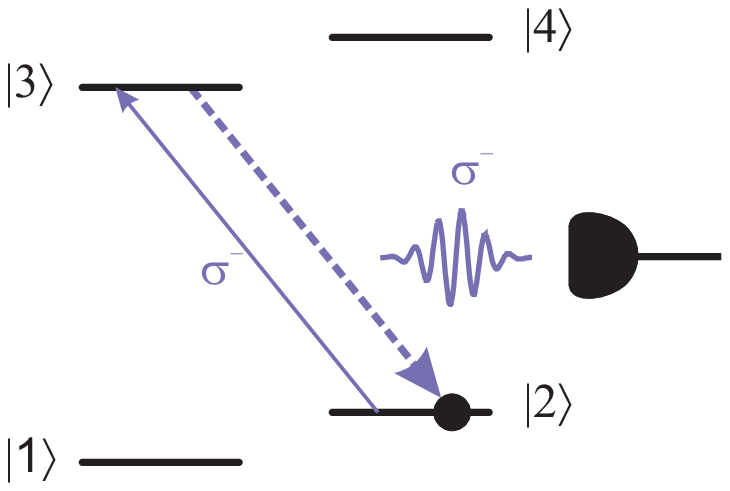}}
\subfloat[$\sigma^{+}$ polarized
photon]{\label{Termschema_Ca_4levels_cond_g2_sigma+}\includegraphics[width=0.650\columnwidth]{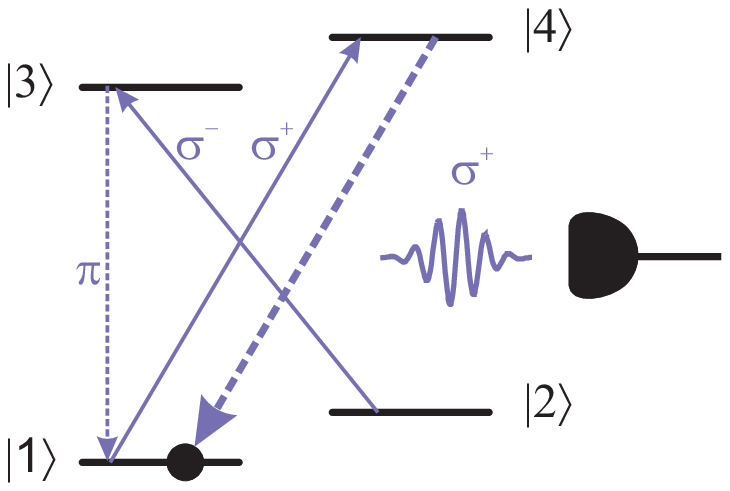}}
 \caption{Detection of $\sigma^{-}$ (b) and $\sigma^{+}$ (c) polarized photons,
 conditioned on the previous detection of a $\sigma^{-}$ (a) polarized photon. Excitation happens with $\sigma^+$-$\sigma^-$-polarized light,
 i.e. linear polarization perpendicular to the magnetic field. (a) Detection of a $\sigma^{-}$ polarized photon projects
the ion into the state $|2\rangle = |S_{1/2}, m_j=1/2\rangle$. (b)
To detect a subsequent $\sigma^{-}$ polarized photon,
 the ion has to be re-excited
   under the absorption of a $\sigma^{-}$ polarized photon. (c) To detect a $\sigma^{+}$ polarized photon, the ion has
to be re-excited under the absorption of a $\sigma^{-}$ polarized
photon, then decay to state $|1\rangle$ under emission of a $\pi$
polarized photon and then be excited to the state $|4\rangle$ by
reabsorbing a $\sigma^+$ polarized photon from the exciting
beams.} \label{conditionedg2}
\end{center}
\end{figure*}
We now consider polarization-selective detection of photons for
the situation of our experiment. Figure \ref{conditionedg2} shows
a sketch of the four levels involved in the emission of blue
(397~nm) photons on the $\mrm{P_{1/2}}$ to $\mrm{S_{1/2}}$
transition. The detection of a $\sigma^-$ polarized photon
projects the ion into state $|2\rangle=|\textrm{S}_{1/2},
m_j=1/2\rangle$ (figure
\ref{Termschema_Ca_4levels_cond_g2_prodject}). If the ion is
continuously illuminated by linearly polarized light at 397 and
866~nm with polarization perpendicular to the magnetic field, no
$\pi$ transitions are excited. The ion effectively sees a
superposition of $\sigma^-$ and $\sigma^+$ polarized light. After
the emission of a $\sigma^-$ photon and the corresponding
projection into state $|2\rangle$, a subsequent excitation can
thus only occur to state $|3\rangle=|\textrm{P}_{1/2},
m_j=-1/2\rangle$ by absorbing a $\sigma^-$ polarized photon. From
state $|3\rangle$ the ion can either decay back to state
$|2\rangle$ under emission of a second $\sigma^-$ polarized photon
(figure \ref{Termschema_Ca_4levels_cond_g2_sigma-}), or it can
decay to state $|1\rangle=|\textrm{S}_{1/2}, m_j=-1/2\rangle$
under emission of a $\pi$ polarized photon. Thus the next emitted
photon after a $\sigma^-$ photon can not be $\sigma^+$ polarized.
In order to emit a $\sigma^+$ polarized photon, the ion must first
be excited to state $|4\rangle=|\textrm{P}_{1/2}, m_j=1/2\rangle$,
which can only happen out of state $|1\rangle$ by absorbing a
$\sigma^+$ polarized photon from the exciting beams (figure
\ref{Termschema_Ca_4levels_cond_g2_sigma+}). The probability of
detecting a $\sigma^+$ polarized photon after the time $\tau$
conditioned on the previous detection of a $\sigma^-$ polarized
photon is therefore much lower than the probability of detecting a
$\sigma^-$ polarized photon for the same condition. Since the
correlation function is a measure for the photon-photon waiting
time distribution, the $g^{(2)}$ derived from correlating
$\sigma^-$ with $\sigma^+$ photons should be suppressed with
respect to the $g^{(2)}$ derived from correlating $\sigma^-$ with
$\sigma^-$ photons. In fact both correlations are expected to
exhibit almost ideal antibunching, with the difference that for
the first case the dip around $\tau$=0 is expected to be much
wider. In other words, and taking into account also the case where
$\sigma^+$ and $\sigma^-$ are exchanged, there will be a much
stronger antibunching for the correlation of photons with opposite
$\sigma$ polarization than for photons with the same $\sigma$
polarization. The same considerations are applicable for
excitation with pure $\pi$-polarized light. In that case the
correlation of subsequent photons is much stronger for opposite
than for equal $\sigma$ polarization.

Analogous to equation \ref{g2rho3344}, the second order
correlation functions for $\sigma^-$ and $\sigma^+$ light,
conditioned on the previous detection of a $\sigma^-$ photon
($\hat{\rho}(0)=\hat{\rho}_{\mrm{init}}=|2\rangle \langle 2|$) are
derived to be
\begin{equation}
g^{(2)}_{\sigma^-}(\tau)
=\frac{\rho_{33}(\tau)}{\rho_{33}(\infty)} \label{gcond5}
\end{equation}
and
\begin{equation}
g^{(2)}_{\sigma^+}(\tau)
=\frac{\rho_{44}(\tau)}{\rho_{44}(\infty)}, \label{gcond6}
\end{equation}
respectively. In contrast to the general case (equation
\ref{g2rho3344}) the conditioned second order correlation function
is given by only one of the excited state populations at time
$\tau$.

\begin{figure*}
\begin{center}
\subfloat[]{\label{g2_theory}\includegraphics[width=0.98\columnwidth]
{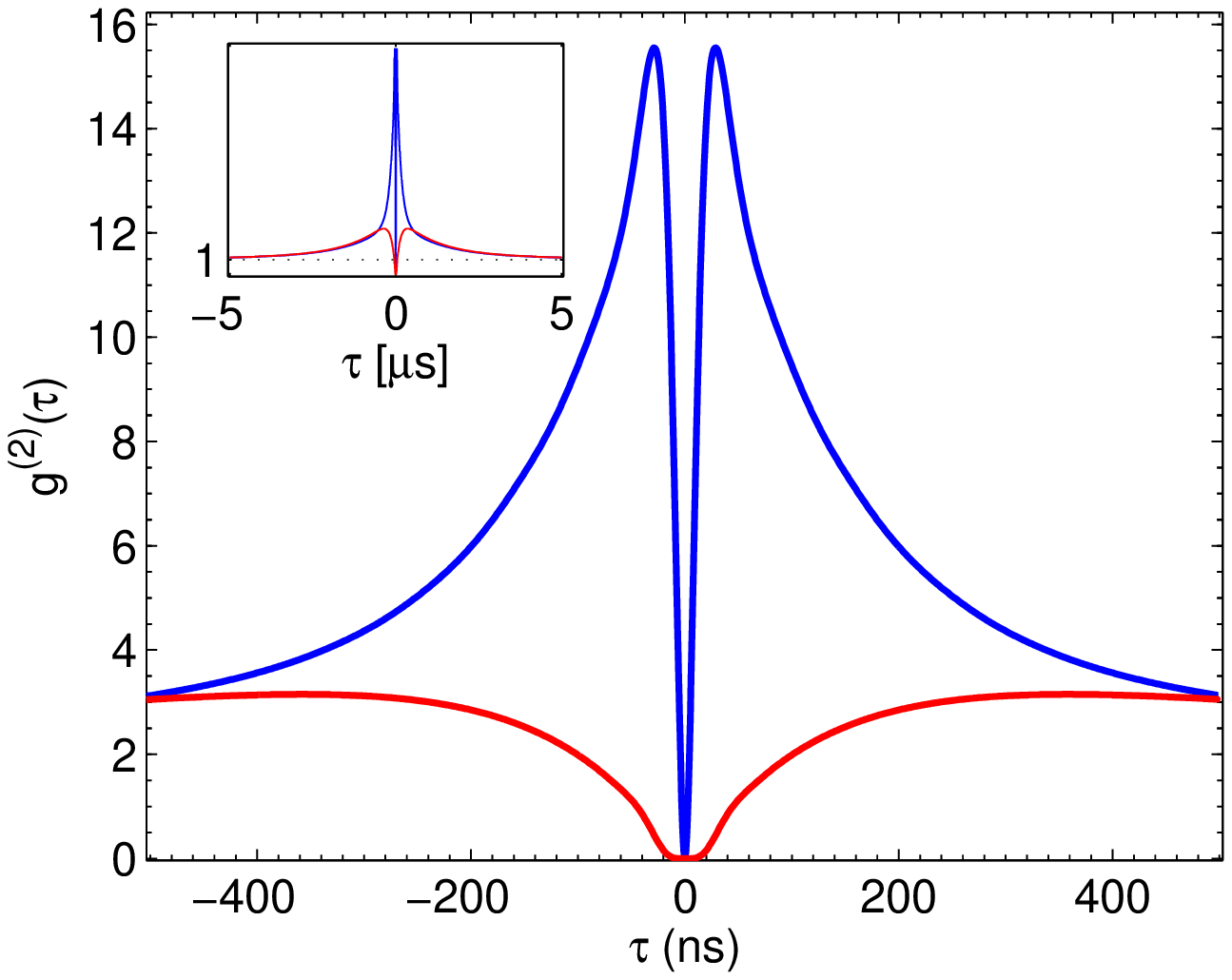}}
 \subfloat[]
   {\label{g2_theory_zoom}\includegraphics[width=0.98\columnwidth]{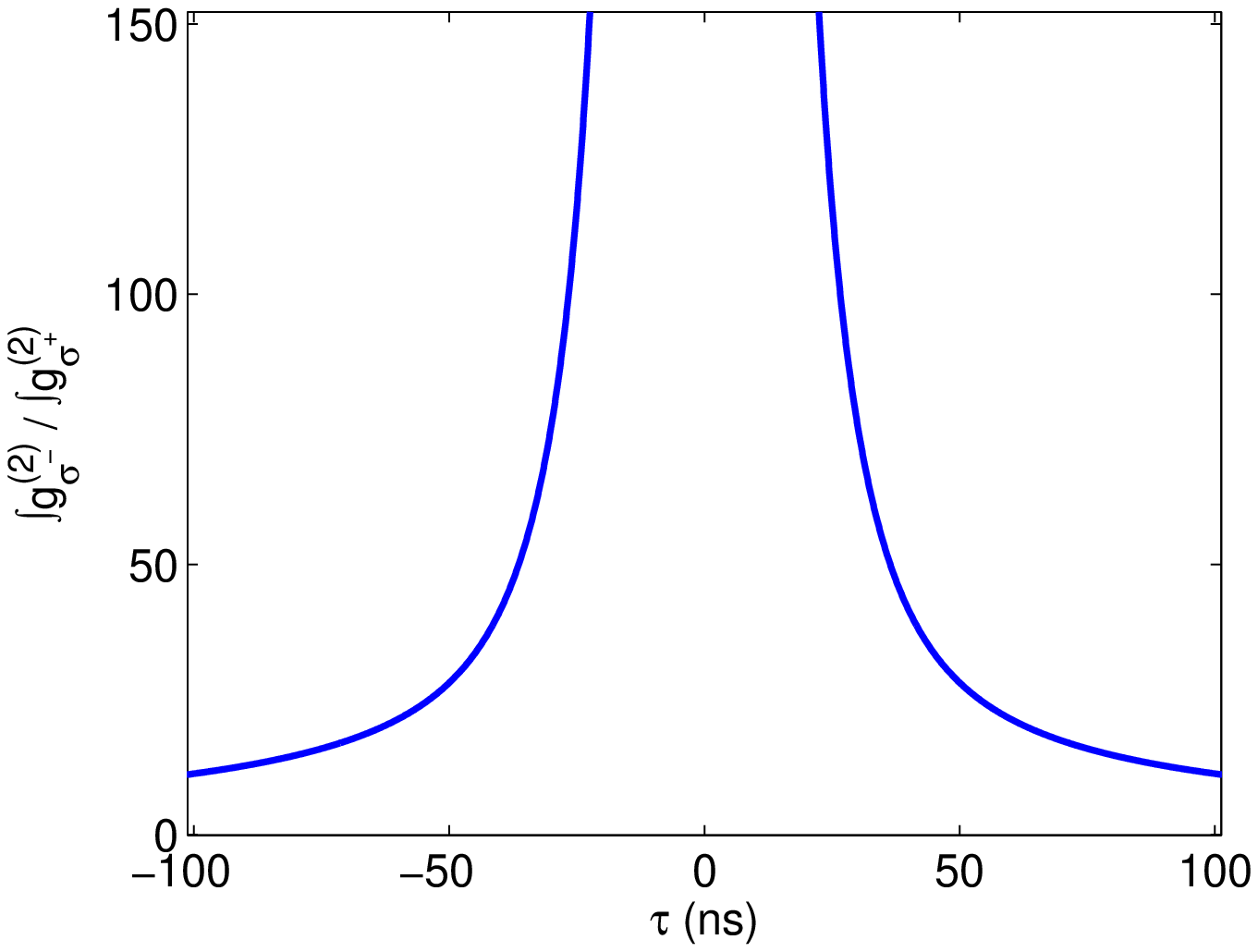}}
 \caption{(a) Conditioned second order correlation functions $g^{(2)}_{\sigma^-}(\tau)$
 (top, blue) and $g^{(2)}_{\sigma^+}(\tau)$ (bottom, red) for weak
excitation. The functions
 have been calculated with a eight level Bloch equation model for Rabi frequencies of $\Omega_{397}=2\pi \cdot 9.2\,\mrm{MHz}$,
 $\Omega_{866}=2\pi \cdot 1.3\,\mrm{MHz}$, detunings $\Delta_{397}/2\pi=-15\,\mrm{MHz}$, $\Delta_{866}/2\pi=5.8\,\mrm{MHz}$ and
 a magnetic field of $\mrm{B}=3.5\,\mrm{G}$. (b) Ratio $p(\tau)$ (equation \ref{ratio}) of the integrals over time of the two
conditioned correlation functions.} \label{g2_theory_fig}
\end{center}
\end{figure*}

\begin{figure*}
\begin{center}
\subfloat[]{\label{g2_theory_strong_excitation}\includegraphics[width=0.98\columnwidth]
{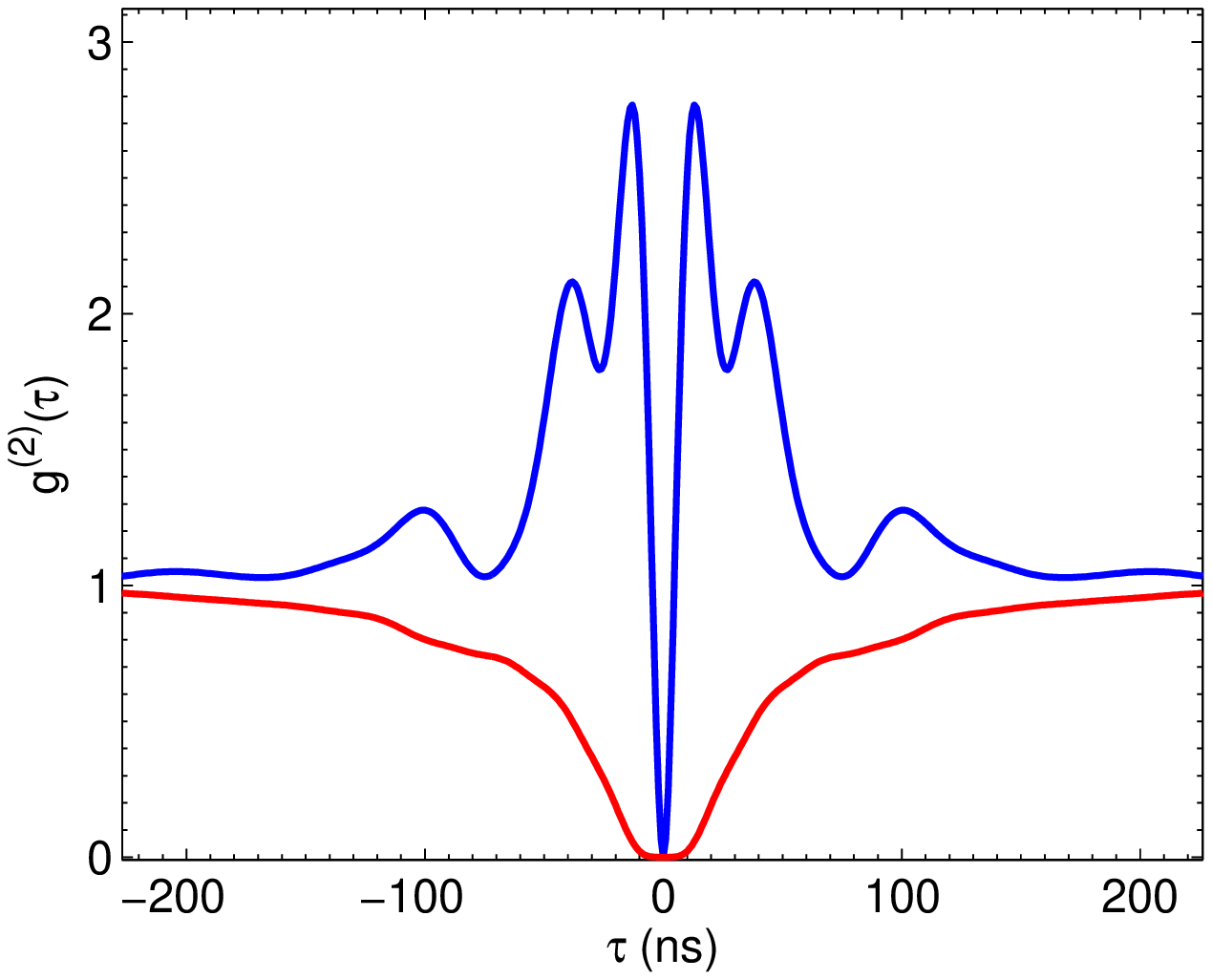}}
 \subfloat[]
   {\label{g2_theory_strong_excitation_zoom}\includegraphics[width=0.98\columnwidth]{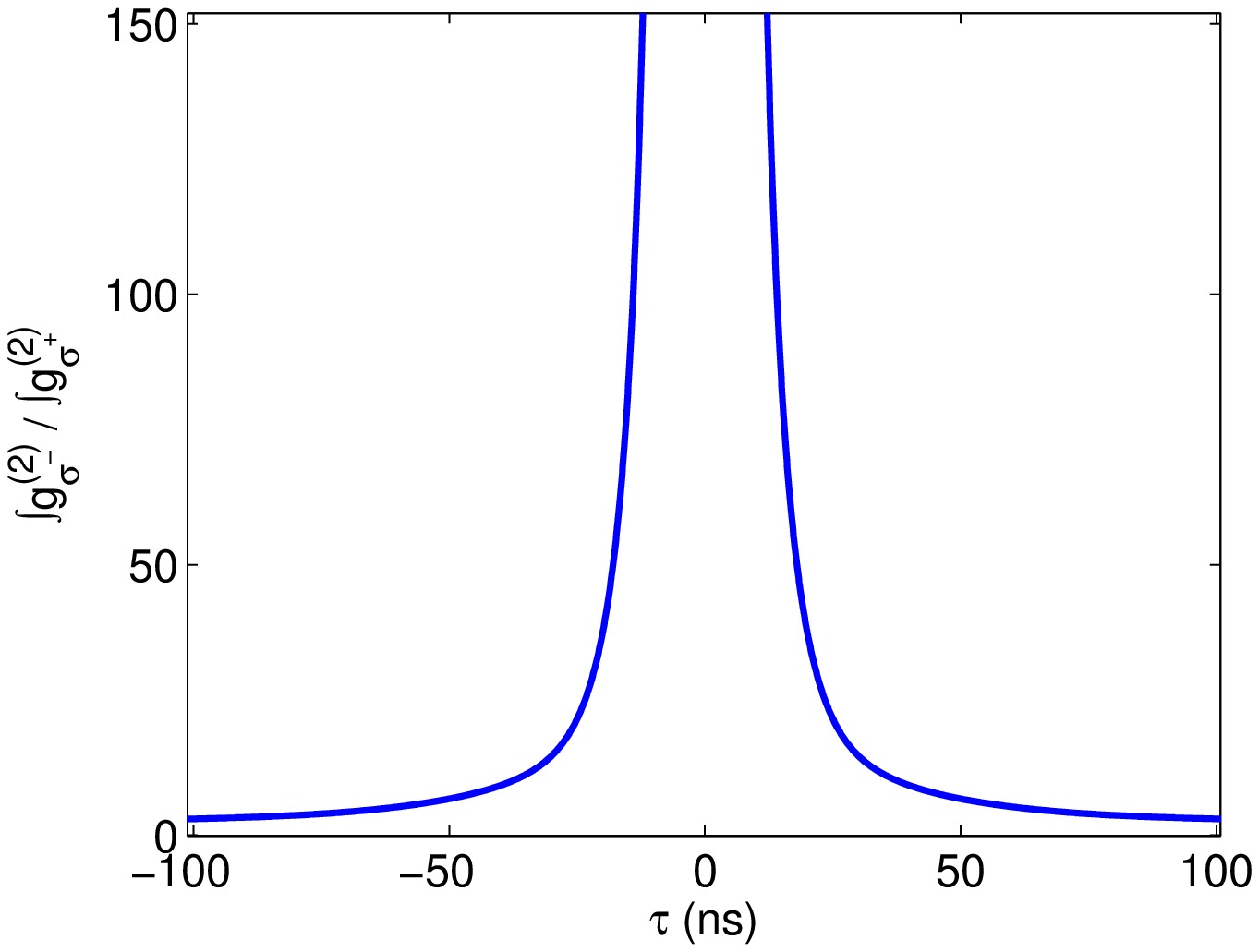}}
 \caption{(a) Conditioned second order correlation functions $g^{(2)}_{\sigma^-}(\tau)$
 (top, blue) and $g^{(2)}_{\sigma^+}(\tau)$ (bottom, red) for
 strong excitation. The functions
 have been calculated with a eight level Bloch equation model for Rabi frequencies of $\Omega_{397}=2\pi \cdot 20.2\,\mrm{MHz}$,
 $\Omega_{866}=2\pi \cdot 20.3\,\mrm{MHz}$, detunings $\Delta_{397}/2\pi=-15\,\mrm{MHz}$, $\Delta_{866}/2\pi=5.8\,\mrm{MHz}$ and
 a magnetic field of $\mrm{B}=3.5\,\mrm{G}$. (b) Ratio $p(\tau)$ (equation \ref{ratio}) of the integrals over time of the two
conditioned correlation functions.}
\label{g2_theory_strong_excitation_fig}
\end{center}
\end{figure*}

In figure \ref{g2_theory} and \ref{g2_theory_strong_excitation}
the correlation functions calculated according to equations
\ref{gcond5} and \ref{gcond6} are plotted for weak and strong
excitation parameters typical for our experimental setup. The
upper curves, in blue, show $g^{(2)}_{\sigma^-}(\tau)$, while the
lower ones, in red, represent $g^{(2)}_{\sigma^+}(\tau)$. As
expected $g^{(2)}_{\sigma^-}(\tau)$ and $g^{(2)}_{\sigma^+}(\tau)$
show very different behaviors. For both weak and strong
excitation, $g^{(2)}_{\sigma^-}(\tau)$ rises to high values with a
steep slope. For weak excitation a maximum $g^{(2)}$ of 15.6 is
reached at $\tau=\pm 29\,\mrm{ns}$. For longer time differences
the correlation function falls monotonously until it reaches one.
For strong excitation $g^{(2)}_{\sigma^-}(\tau)$ reaches a maximal
value of 2.8 at $\tau=\pm 13\,\mrm{ns}$. For $\tau > 13\,\mrm{ns}$
the correlation function falls to 1 after 200~ns showing some
coherent oscillations.

The two $g^{(2)}_{\sigma^+}(\tau)$ functions, on the contrary,
show a flat behavior on short time scales before they rise with a
moderate slope to their maximum value, which is reached at
approximately 400~ns for the weak excitation and 200~ns for the
strong excitation. In the latter case $g^{(2)}_{\sigma^+}(\tau)$
rises directly to 1 without a transient overshoot, while for weak
excitation it reaches a value of 3.1 before it falls to 1 for
$\tau > 400\,\mrm{ns}$.

The shape of $g^{(2)}_{\sigma^-}(\tau)$ in figure \ref{g2_theory}
is characterized by large correlation values and a long decay time
(compared to the lifetime of the $\mrm{P_{1/2}}$ state) to the
asymptotic value and can be attributed to optical pumping into the
$\mrm{D_{3/2}}$ state \cite{Schubert1995PRAv52p2994}. The
excitation on the $\mrm{S_{1/2} - P_{1/2}}$ transitions is much
stronger than the one on the $\mrm{D_{3/2} - P_{1/2}}$
transitions. As a result, a large fraction of the population is
transferred to the $\mrm{D_{3/2}}$ state after long delay times
and the observed fluorescence is weak. This small flux of
fluorescence, caused by a small steady-state population of the
state $|3\rangle$, determines the normalization factor for long
time intervals $\rho_{33}(\infty)$. At short time delays after the
projection into state $|2\rangle$, however, i.e. during the first
30-40~ns, a large fraction of the population is excited to state
$|3\rangle$ and the optical pumping to the $\mrm{D_{3/2}}$ states
is negligible. Since the correlation function is the ratio of the
population of state $|3\rangle$ at time $\tau$ and that in the
steady state, high values are reached, which then decay to one
revealing the time scale of the optical pumping.

The characteristics of $g^{(2)}_{\sigma^+}(\tau)$ can be explained
with an analogous argumentation. Here, excitation to state
$|4\rangle$ during the first 30-40~ns is even smaller than the
steady state population $\rho_{44}(\infty)$ for long time delays.
After the projection into $|2\rangle$, it takes several scattering
events and therefore more time until the population of state
$|4\rangle$ exceeds $\rho_{44}(\infty)$. As the inset of figure
\ref{g2_theory} shows, for large time intervals
$g^{(2)}_{\sigma^+}(\tau)$ decays to the asymptotic value with the
same time constant as $g^{(2)}_{\sigma^-}(\tau)$.

The correlation functions shown in figure
\ref{g2_theory_strong_excitation} are calculated for equal
excitation strength on the $\mrm{S_{1/2} - P_{1/2}}$ and
$\mrm{D_{3/2} - P_{1/2}}$ transitions. Consequently, the
correlation values of $g^{(2)}_{\sigma^-}(\tau)$ are smaller and
the decay to the asymptotic value is much faster. The exciting
fields are strong enough to cause some damped oscillations at the
generalized Rabi frequency
$\Omega_G=\sqrt{|\Omega_{397}|^2+\Delta_{397}^2}$ in the
correlation function.
Due to the complex eight-level structure, the oscillations do not
occur at only one generalized Rabi frequency, but each transition
between the Zeeman sub-levels contributes a Fourier component
depending on the intensity and detuning of the driving field. In
figure \ref{g2_theory_strong_excitation} this becomes noticeable
by comparing the frequency of the strongly suppressed oscillations
of $g^{(2)}_{\sigma^+}(\tau)$ with the ones from
$g^{(2)}_{\sigma^-}(\tau)$. Due to the Zeeman splitting, the
detunings of the $|1\rangle$ to $|4\rangle$ and the $|2\rangle$ to
$|3\rangle$ transitions with respect to the exciting laser give
rise to different generalized Rabi frequencies.

The most interesting feature in figure \ref{g2_theory} and
\ref{g2_theory_strong_excitation} is the behavior of the
correlation function for times $\tau$ close to zero.
$g^{(2)}_{\sigma^+}(\tau)$ shows a flat plateau of values very
close to zero for $-15\,\mrm{ns}<\tau<15\,\mrm{ns}$ in the case of
weak and for $-10\,\mrm{ns}<\tau<10\,\mrm{ns}$ in the case of
strong excitation. In the same time interval
$g^{(2)}_{\sigma^-}(\tau)$ rises very fast to high values for both
excitation conditions. The ratio
\begin{equation}
p(\tau)=\frac{\int_0^{\tau} g^{(2)}_{\sigma^-}(t)dt}{\int_0^{\tau}
g^{(2)}_{\sigma^+}(t)dt} \label{ratio}
\end{equation}
of the number of $\sigma^-$- and $\sigma^+$-polarized photons
emitted by the ion in $\tau$ describes the purity of the
polarization in that time window. The probability with which, if a
second photon is detected within this time interval, the
polarization of this second photon is $\sigma^-$, is easily
calculated as $\frac{p}{1+p}$. Figure \ref{g2_theory_zoom} and
\ref{g2_theory_strong_excitation_zoom} show $p$ for weak and
strong excitation conditions, respectively. The ratio diverges for
$\tau \rightarrow 0$ ($>10^5$ for $\tau=1\,\mrm{ns}$) in both
excitation regimes, because $\rho_{33}(\tau)$ increases
quadratically in time from $\tau=0$, while $\rho_{44}(\tau)$
increases with $\tau^4$. In other words, if a second photon is
emitted within a short time interval after a first
$\sigma^-$-polarized photon, then with very high probability (that
can be chosen by the time interval) it will also be
$\sigma^-$-polarized. By exciting the ion with $\pi$-polarized
light it is possible to generate equally strongly correlated
photon pairs of oppositely $\sigma$-polarized photons.

\section{Experiment} \label{Experiment}
\begin{figure}
\begin{center}
\includegraphics[width=\columnwidth]{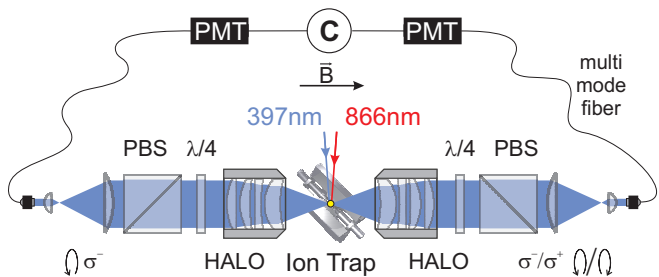}
\caption{Setup for the measurement of polarization-conditioned
correlation functions. The fluorescence light is split in to two
parts by collecting it with the two HALO lenses. Multimode fibers
are used to couple the light to two PMTs.} \label{setup_g2}
\end{center}
\end{figure}
Figure \ref{setup_g2} shows the setup of the measurement which is
described in detail elsewhere
\cite{Gerber2009NJoPv11p13,Rohde2009a[vp}. A single
$^{40}\mrm{Ca^+}$ ion is trapped in a linear Paul trap and
continuously excited by two lasers at $397$~nm and $866$~nm. The
$397\,\mathrm{nm}$ laser is red detuned to provide Doppler cooling
while the resonant $866$~nm laser prevents optical pumping to the
$\mrm{D_{3/2}}$ state. The 397~nm fluorescence light is split into
two parts by collecting it with two high numerical aperture lenses
(HALO) \cite{Gerber2009NJoPv11p13}. Both lenses direct the
collected photons to photo multipliers (PMT) through multimode
fibers. In each beam path a $\lambda/4$ plate and a polarizing
beam splitter are used to select the respective polarization. The
arrival times of the signals from the PMTs are recorded with
picosecond resolution by commercial counting electronics
\footnote{Pico Harp 300, Pico Quant}, and the correlation function
is obtained by postprocessing the data.

\subsubsection{Calibration} 
\begin{figure}
\begin{center}
\includegraphics[width=\columnwidth]{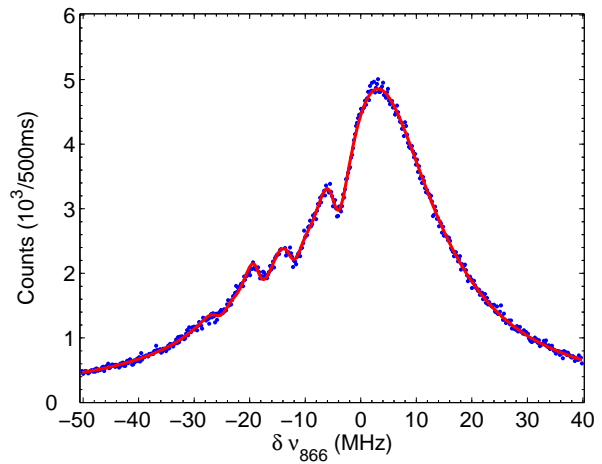}
\caption{Excitation spectrum of a single ion. The 397 and 866~nm
lasers are approximately vertically polarized and propagate under
$90^{\circ}$ to the quantization axis (see figure \ref{setup_g2}).
The solid line (red) is the calculated spectrum for the Rabi
frequencies $\Omega_{397}=2\pi \cdot 9.9\,\mrm{MHz}$,
$\Omega_{866}=2\pi \cdot 1.5\,\mrm{MHz}$, the detuning
$\Delta_{397}/2\pi=-15\,\mrm{MHz}$ and a magnetic field of
$\mrm{B}=3.5\,\mrm{G}$. The background is 89 c.p.s. and the data
points shown have Poissonian errors.}
\label{fit20090320_1825_GPC2e}
\end{center}
\end{figure}

Before measurement of the
correlation functions, an excitation spectrum, i.e. the rate of
397~nm fluorescence as a function of the detuning of 866~nm, was
recorded. The spectrum that shows four dark resonances provides a
calibration of the experimental parameters, which are then used as
the starting point to fit the conditioned correlation functions.
In figure \ref{fit20090320_1825_GPC2e} this excitation spectrum is
plotted.
The Rabi frequencies, detunings and the magnetic field which are
deduced from a fit to the data are indicated in the caption. To
account for experimental deviations in the polarization angle of
the excitation lasers from the ideal vertical polarization, this
parameter was also varied in the fit. The angles of the
polarizations of the two laser beams with the magnetic field, as
used in the model shown in figure \ref{fit20090320_1825_GPC2e},
are $\alpha_{397}=0.46\cdot \pi$ for the blue laser and
$\alpha_{866}=0.4\cdot \pi$ for the infrared laser (rather than
$\pi/2$ in the ideal case). This is compatible with the available
control over the experimental settings and the quality of the
optical components used.

\subsubsection{Conditioned correlation functions}
\begin{figure}
\begin{center}
\includegraphics[width=\columnwidth]{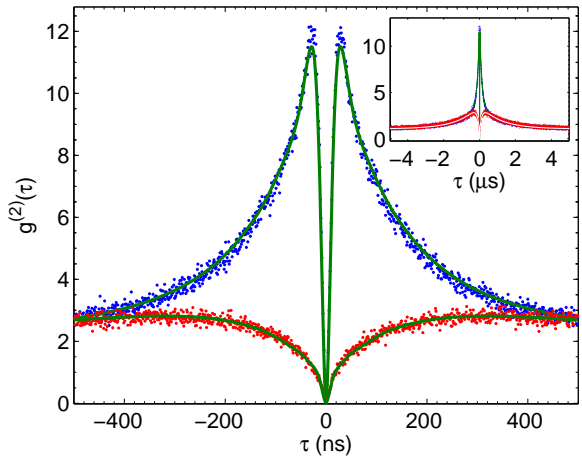}
\caption{Blue (top): ($g^{(2)}_{\sigma^-}$). Red (bottom):
($g^{(2)}_{\sigma^+}$). The solid lines (green) are the
correlation functions expected from the theoretical model for the
Rabi frequencies $\Omega_{397}=2\pi \cdot 9.2\,\mrm{MHz}$,
$\Omega_{866}=2\pi \cdot 1.3\,\mrm{MHz}$ and the detuning
$\Delta_{866}/2\pi=5.8\,\mrm{MHz}$. The inset shows data and model
for time scales up to $5\,\mrm{\mu s}$.}
\label{doublefit_sse_5_EB4e5_2}
\end{center}
\end{figure}
Figure \ref{doublefit_sse_5_EB4e5_2} shows the data for the two
measured conditioned correlation functions plotted in one graph.
Data for $g^{(2)}_{\sigma^+}$ are presented in red (lower curve)
and data for $g^{(2)}_{\sigma^-}$ in blue (upper curve). The data
are normalized to a long-time value of one and presented without
background subtraction. The solid lines are the $g^{(2)}$
functions obtained by a fit of the Rabi frequencies to the
experimental data using the model discussed in section
\ref{Model}. The values of the background, laser detunings,
magnetic field as well as the polarization angles extracted from
the fit to the excitation spectrum have been kept fixed. The two
Rabi frequencies have been fitted to both curves at the same time
and agree well with the experiment. The deviation of the Rabi
frequencies between the fitted correlation functions and the
fitted excitation spectrum lies within the statistical error
\footnote{Since the parameters used in fitting the spectrum as
well as the $g^{(2)}$ functions are not fully independent from
each other, various sets of parameters are consistent with the
data, in the sense that the whole set of parameters for the
spectrum fits the correlations within 1 of the $\chi^2$ deviation
and vice versa.}. The correlation functions are very similar to
the ones for the case of weak excitation in figure
\ref{g2_theory_fig}. For large $\tau$ ($>$ 400~ns) both functions
overlap and slowly decay to one. The characteristic behavior of
large (small) correlation values and the slow decay to the
asymptotic value for $g^{(2)}_{\sigma^-}$ ($g^{(2)}_{\sigma^+}$),
explained in section \ref{Model}, are clearly observed in the
measurement.
\begin{figure}[t]
\begin{center}
\includegraphics[width=\columnwidth]{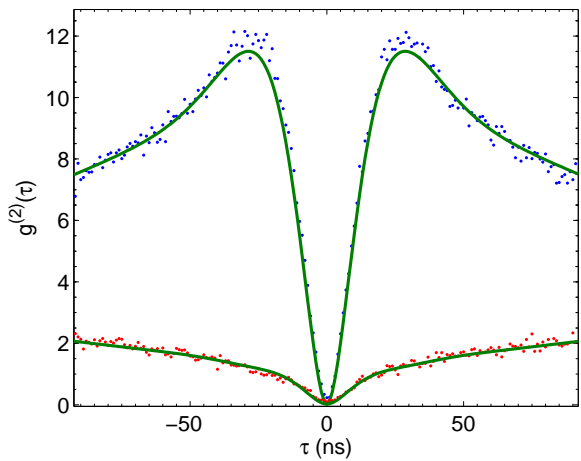}
\caption{Zoom of FIG. \ref{doublefit_sse_5_EB4e5_2}}
\label{g2_fit20090320_1825_GPC2e_zoom}
\end{center}
\end{figure}

Figure \ref{g2_fit20090320_1825_GPC2e_zoom} shows a zoom into the
region for small time differences. At $\tau=0$ both curves reach a
value close to zero. With increasing $\tau$, $g^{(2)}_{\sigma^-}$
rises with a very steep quadratic slope and reaches a value as
large as 12, whereas $g^{(2)}_{\sigma^+}$ stays flat for
$\sim5\,\mrm{ns}$, before it rises with a moderate slope to a
maximum value of 2.9. The model agrees very well with the data,
proving the good control over the creation of polarization
correlated photon pairs that we obtained experimentally.

The main difference between the model calculation of $g^{(2)}$ in
figure \ref{g2_theory} and experimental data in figure
\ref{g2_fit20090320_1825_GPC2e_zoom} is that the predicted
$\tau^4$-like plateau of $g^{(2)}_{\sigma^+}(\tau)$ is less
pronounced in the measurement. Simulations with the model from
section \ref{Model} show that this is explained by small errors in
the polarization of the exciting lasers and in the detection
setup. The polarizations of the exciting lasers have been fitted
to the excitation spectrum (fig \ref{fit20090320_1825_GPC2e}), and
these results have been used in the model of the correlation
functions. To achieve an agreement of the quality as it is shown
in figure \ref{doublefit_sse_5_EB4e5_2} and
\ref{g2_fit20090320_1825_GPC2e_zoom}, also deviations from the
ideal polarizations in the detection were accounted for in the
model. These deviations occur when the polarization is not
filtered perfectly, and consequently the measured $g^{(2)}$
function contain some coincidences that are caused by the
respective orthogonal polarization. The theoretical model in
figures \ref{doublefit_sse_5_EB4e5_2} and
\ref{g2_fit20090320_1825_GPC2e_zoom} has been calculated using
2.5~\% of wrong $\sigma^+$ polarized photons for the initial
detection events in both curves. For $g^{(2)}_{\sigma^-}$ the
conditioned detection of the second $\sigma^-$ polarized photon
has an error of 5\%, while for $g^{(2)}_{\sigma^+}$ the detection
of the $\sigma^+$ polarized photon has an error of 1.8\%. This
demonstrates that $g^{(2)}_{\sigma^+}$, in particular its $\tau^4$
characteristic, is very sensitive to small polarization errors at
times $\tau$ close to zero. These errors are within the precision
with which the polarization filtering was controlled in the
experiment, given the low light levels and imperfections of the
optics. Furthermore, there is a contribution due to the collection
of light from a solid angle along the quantization axis. Each HALO
lens collects 4\% of the light emitted into the full solid angle
\cite{Gerber2009NJoPv11p13}, which leads to the detection of a
small fraction wrong $\sigma$ and $\pi$ polarized photons.

\begin{figure}
\begin{center}
\includegraphics[width=\columnwidth]{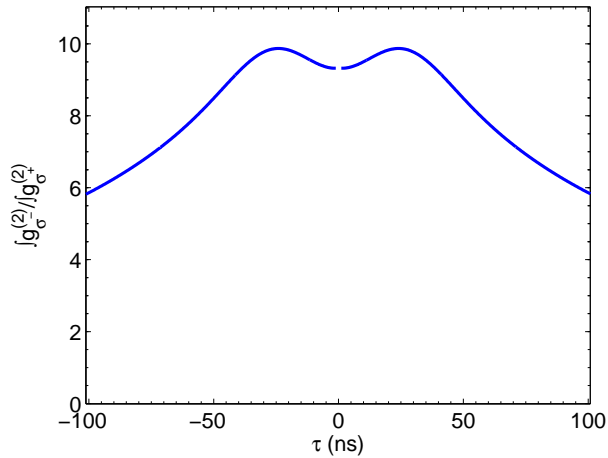}
\caption{Purity $p$ for the model calculation fitted to the
measured data from figure \ref{doublefit_sse_5_EB4e5_2} and
\ref{g2_fit20090320_1825_GPC2e_zoom}. To avoid artifacts at times
close to zero $p$ is presented without background.}
\label{integral_ratio_GPC2e_no_background_paper}
\end{center}
\end{figure}

The sensitivity of the generation of polarization correlated
photon pairs to polarization errors gets even more evident
considering the purity $p$ (equation \ref{ratio}). Figure
\ref{integral_ratio_GPC2e_no_background_paper} shows this ratio
for the values of the model calculation that was fitted to the
data from figure \ref{doublefit_sse_5_EB4e5_2} and
\ref{g2_fit20090320_1825_GPC2e_zoom}.
In contrast to the ideal case from figure \ref{g2_theory_zoom},
$p$ does not diverge for times $\tau \rightarrow 0$, but instead
reaches a maximum of almost 10 at 24~ns and falls then to 9.3 at
1~ns. This means, if a second photon is detected with our setup
within 24~ns, it is $\sigma^-$ polarized with 91~\% probability. A
practical figure of merit also has to consider the absolute number
of photons within the time interval $\tau$. After 24~ns the ion
emits on average 0.07 $\sigma^-$ polarized photons into the full
solid angle, out of which 8\% are detected with our setup
\cite{Gerber2009NJoPv11p13}. Increasing the time window will yield
more photons, but decrease the polarization purity. The curve for
the ideal case in figure \ref{g2_theory_zoom} reaches a value of
130 at 24~ns, suggesting that it is in principle possible to
generate polarization correlated photon pairs with more than 99\%
probability using this method. This could be achieved by reducing
polarization errors in excitation and detection.

The excitation conditions of the single ion also affect the
efficiency of our photon pair source. For the strong excitation
conditions of figure \ref{g2_theory_strong_excitation_fig}, for
example, the ion emits on average 0.2 $\sigma^-$ polarized photons
within 24~ns. Simultaneously the purity decreases, yielding a
$\sigma^-$ polarized photon in only 96\% of the cases (figure
\ref{g2_theory_strong_excitation_zoom}). For a good comparison of
the source under weak and strong excitation conditions it is
convenient to look at the efficiency of both cases for equal
photon numbers. Under strong excitation conditions the ion emits
on average 0.07 photons in 12~ns. If a second photon is detected
within these 12~ns, it is $\sigma^-$ polarized with more than
99\%. Increasing the Rabi frequencies of the excitation beams thus
allows to reduce the time interval in which the photon pairs are
emitted while keeping photon number and polarization purity
constant.

Summarizing, it was shown that the second order correlation
function of the fluorescence light of a single ion can be
engineered by polarization-sensitive detection. The $g^{(2)}$
functions for $\sigma^+$ and $\sigma^-$ light conditioned on the
previous detection of a $\sigma^-$ photon show a characteristic
anti-bunching behavior that allows for heralding photons of a
certain polarization in an otherwise randomly polarized stream of
photons, with possible applications in quantum optical information
technology. A single, laser cooled ion generated polarization
correlated photon pairs within a time window of 24~ns with an
efficiency of 91\%. Model calculations show that the presented
method has the potential to reach an efficiency of more than 99\%
within the same time window.

We thank Giovanna Morigi for helpful discussions.

We acknowledge support from the European Commission (SCALA,
contract 015714; EMALI, MRTN-CT-2006-035369), the Spanish MICINN
(QOIT, CSD2006-00019; QLIQS, FIS2005-08257; QNLP, FIS2007-66944),
and the Generalitat de Catalunya (2005SGR00189; FI-AGAUR
fellowship of C.S.).

\end{document}